\documentclass[12pt]{article}
\usepackage{amssymb}
\begin{document}

\def\ad{\mbox{ad}\,}
\def\ker{\mbox{Ker}\,}   
\def\det{\mbox{det}\,}

\def\vev#1{\langle #1 \rangle}
\def\bra#1{\langle #1 \vert}
\def\ket#1{\vert #1 \rangle}


\newcommand{\sect}[1]{\setcounter{equation}{0}\section{#1}}
\renewcommand{\theequation}{\thesection.\arabic{equation}}
\newcommand{\subsect}[1]{\setcounter{equation}{0}\subsection{#1}}
\renewcommand{\theequation}{\thesubsection.\arabic{equation}}

\newcommand{\be}{\begin{equation}}
\newcommand{\ee}{\end{equation}}
\newcommand{\bea}{\begin{eqnarray}}
\newcommand{\eea}{\end{eqnarray}}

\newcommand{\px}{\partial_{x}}
\newcommand{\py}{\partial_{y}}
\newcommand{\pu}{\partial_{u}}
\newcommand{\pv}{\partial_{v}}

\title{ Quantum Mechanics  on the $h$-deformed \\ Quantum Plane}
\author{Sunggoo Cho\\ Department of Physics, Semyung University, \\
        Chechon, Chungbuk 390 - 711, Korea
           }

%

\maketitle

\begin{abstract}

We find the covariant deformed Heisenberg algebra and the Laplace-Beltrami
operator on the extended $h$-deformed quantum plane 
and solve the
Schr\"odinger equations explicitly for some  physical systems  on the quantum
plane. 
In the commutative limit the behaviour of a quantum particle on
the quantum plane becomes that of the quantum particle on the
Poincar\'e half-plane, a surface of constant negative Gaussian
curvature.  We show  
the bound state energy spectra for particles under specific potentials
 depend explicitly on the deformation parameter $h$.
Moreover,   
it is shown that bound states can survive on the quantum plane
 in a limiting case where 
bound states on the Poincar\'e half-plane disappear. 

\end{abstract}



\vfill








\newpage

\parskip 4pt plus2pt minus2pt

\sect{Introduction }

In recent years, quantum groups and quantum spaces have attracted much
attention  from both  physics and mathematics~\cite{Con94, Mad95, Man88}. 
They are of relevance to the problem of the quantization of spacetime.
Quantum   spaces 
are often adopted as models for the microscopic structure of
  physical spacetime  and their effects on  physics have been
investigated
by many authors.
   In particular, $q$-deformed quantum spaces are one of those quantum spaces
whose effects have been intensively studied    
 (see, e.g. \cite{AreVol91, FicLorWes96, LorRufWes97}). 
                                             
  The purpose of this work is to make a formulation of 
quantum mechanics on the $h$-deformed quantum 
plane and to investigate the effects of the quantum plane on the energy spectra.
The $h$-deformed quantum plane~\cite{Agh93, DemManMukZhd90, Kup92}  is a counterpart of the $q$-deformed 
one in the set of quantum planes which are covariant under those
quantum deformations of $GL(2)$ which admit a central determinant.
It seems that it has more geometrical structures than the $q$-deformed
one. The $h$-deformed quantum plane is
known to be  a noncommutative version of the Poincar\'e half-plane 
and its  geometry  has  been
  discussed in \cite{ChoMadPar98}.

In Section 2, we review the geometry of the $h$-deformed
quantum plane which is to be used later.  In Section 3, the
$h$-deformed Heisenberg algebra is constructed from the skew
derivatives of Wess-Zumino. 
This algebra is covariant under the
quantum group $GL_{h}(2)$. It is worthy of being compared with 
the deformed Heisenberg algebra used by Aghamohammadi~\cite{Agh93}, 
 which is not covariant under the quantum group.
Also it is comparable with
 the deformed Heisenberg algebra,
        for example, by Lorek {\it et
al.}~\cite{LorRufWes97}, which has no such quantum group symmetry
because there is none compatible with the reality condition. 
We also
construct the Laplace-Beltrami operator on the quantum plane. The
operator has the Laplace-Betrami operator of the Poincar\'{e}
half-plane as its commutative limit. 
 In Section 4, we  construct
the Schr\"{o}dinger equations for some physical systems  
 on
the quantum plane and find their solutions explicitly, taking values in 
the noncommutative algebra. 
 In the commutative limit the behaviour of a quantum particle on
the quantum plane becomes that of the quantum particle on the
Poincar\'e half-plane, a surface of constant negative Gaussian
curvature. 
The bound state energy spectra for particles under specific potentials are shown to
 depend explicitly on the deformation parameter $h$.  
This result is comparable with that of \cite{CerHinMadWes98}.
  Moreover,   
it is shown that bound states can survive on the quantum plane
 in a limiting case where 
bound states on the Poincar\'e half-plane disappear.

\sect{The $h$-deformed quantum plane }

\subsect{ The covariant differential structure }

The $h$-deformed quantum plane is an associative
algebra 
generated by noncommuting coordinates
$x$ and $y$ such that
\be
xy - yx = h y^2,          \label{2.1}
\ee
where $h$ is a deformation parameter.  The quantum group $GL_{h}(2)$ is
the symmetry group of the $h$-deformed plane as is $GL_{q}(2)$ for the
$q$-deformed quantum plane~\cite{ChaPre94,DemManMukZhd90,Kup92}.

The covariant differential calculus on the quantum plane can be
found~\cite{Agh93} by the method of Wess and Zumino~\cite{WesZum90}.
The results to be used in this work can be summarized as follows.
The module structure of the 1-forms is given by the relations
\be 
\begin{array}{ll}
x dx = dx \, x - h dx \, y +h dy \, x + h^2 dy \, y,
&x dy = dy \, x +h dy \, y, \\
y \, dx = dx \, y - h dy \, y,
&y dy = dy \, y,
\end{array}                                                     \label{2.3}
\ee
and the structure of the algebra of forms is determined by the relations
\be
dx^2 =  h dx dy,
\qquad dx dy = - dy dx,  \qquad
dy^2 = 0.                                                      \label{2.4}
\ee
It is important to notice that the associative algebra
generated by $x, y$ has an
involution given by $x^{\dagger} = x$, $y^{\dagger} = y$
 provided $h \in i \mathbb{R}$.
 The involution has a natural and simple extension
to the differential calculus and  the differential is real:
\be
(df)^{\dagger} = df^{\dagger}.
\ee
This is contrary to the case considered by Lorek {\it et
al.}~\cite{LorRufWes97}. Dual to these forms are a set of twisted
derivations which when considered as operators satisfy the relations
\be
\begin{array}{ll}
 \, [x, \partial_{x}] = -1 + hy\partial_{x}, 
     & [x, \partial_{y}] =  -hx\partial_{x} - h^2 y\partial_{x}
                  - h y\partial_{y}, \nonumber \\
 \,  [y, \partial_{x}] = 0,
      & [y, \partial_{y}] = -1 + hy\partial_{x}, \\
 \,  [\partial_{x}, \partial_{y}] =  h\partial^{2}_{x}.
 & \nonumber
\end{array}                                                  
 \label{2.5}
\ee
In \cite{Agh93} the momentum operators are defined  as follows
\be
p_{x} = -i\hbar \partial_{x}, \qquad  p_{y} = -i\hbar \partial_{y}.
\label{2.50}
\ee
However, if one requires that  they be Hermitian,
the deformed Heisenberg algebra given by (\ref{2.5})
 does not satisfy the hermiticity. This problem of nonhermiticity
has been  observed by
Aghamohammadi~\cite{Agh93}. We shall resolve it in Subsection 3.1. 

The extended $h$-deformed quantum plane is an associative algebra
$\cal{A}$ generated by $x$, $y$ satisfying Equation~(\ref{2.1}) and 
their inverses $x^{-1}$, $y^{-1}$. The extended $h$-deformed quantum plane is
known to be a noncommutative version of a Poincar\'e
half-plane~\cite{ChoMadPar98}. 
 Since $\cal{A}$ is a unital involution algebra
with $x$ and $y$ Hermitian and $h \in i \mathbb{R}$, the elements
\be
u = x y^{-1} + {1\over 2} h, \qquad v = y^{-2}             \label{2.6}
\ee
are also Hermitian. Their commutation relation becomes
\be
[u,v] = - 2 h v.                                                \label{2.7}
\ee
This choice of generators is useful in studying the commutative limit.

Now it is straightforward to see that
\be
\begin{array}{ll}
 u du = du \, u -2hdu, & u dv = dv \, u - 2h dv, \\
 v du = du \, v,        &  v dv = dv \, v, 
\end{array} \label{2.8}
\ee
and
\be
\begin{array}{ll}
du du = dv dv = 0, & du dv = - dv du.  
\end{array} \label{2.9}
\ee
In this work, however, it is more convenient to use the
`Stehbein'~\cite{MadMou98}.

\subsect{ The Stehbein }

The Stehbein $\theta^a$ are defined as
\be
\theta^1 = v^{-1} du, \qquad \theta^2 = - v^{-1} dv.     \label{2.10}
\ee
Then the $\theta^a$ satisfy the commutation relations
\be 
f \theta^a = \theta^a f   \label{2.11}
\ee
for any $f \in \cal{A} $
 as well
as the relations
\be
(\theta^1)^2 = 0, \qquad (\theta^2)^2 = 0, \qquad
\theta^1 \theta^2 + \theta^2 \theta^1 = 0.       \label{2.12}
\ee
Also we have
\be
d\theta^1 = -\theta^1\theta^2, \qquad d\theta^2 = 0.
\label{2.13}
\ee
Moreover, the Stehbein satisfy
\be
\theta^a (e_b ) = \delta^a_{b}
\label{2.14}
\ee
if we introduce the derivations $e_a = \ad \lambda_a$ with
\be
\lambda_1 = {1 \over 2h} v, \qquad
\lambda_2 = {1 \over 2h} u.        
   \label{2.15}
\ee
It is easy to see that the derivations satisfy
\be
\begin{array}{ll}
e_1 u = v,   &e_1 v = 0,  \\
e_2 u = 0,   &e_2 v = - v.    \\
\end{array}               
 \label{2.16}
\ee
The derivations $e_a$ define, in the commutative limit,  vector fields
\be
X_a = \lim_{h \rightarrow 0} e_a,                    
\label{2.17}
\ee
with
\be
X_1 = \tilde v \partial_{\tilde u}, \qquad
X_2 = - \tilde v \partial_{\tilde v},
\label{2.18}
\ee                                  
where $\tilde{u}$,  $\tilde{v} $   
are the commutative
limits of the generators $u$, $v$ of the algebra
$\cal{A}$. 
 The algebra
$\cal{A}$ with the differential calculus defined by the relations
(\ref{2.12}) can be regarded as a noncommutative deformation of the
Poincar\'e half-plane~\cite{ChoMadPar98}. 
In this case, a metric is defined on the tensor product of the 
${\cal A}$-module of 1-forms by 
\be
g(\theta^a \otimes \theta^b ) = \delta^{ab}.
\label{metric}
\ee
The metric satisfies 
\be
g(du \otimes du) = g(dv \otimes dv ) = v^2 ,
\qquad 
g(du \otimes dv) = g(dv \otimes du ) = 0.
\ee
 In terms of the commutative limit $\tilde \theta^a$ of
the Stehbein $\theta^a $, the metric 
 is given in the commutative limit by the line element 
\be
ds^2 = (\tilde \theta^1)^2 + (\tilde \theta^2)^2 =
\tilde v^{-2} (d \tilde u^2 + d \tilde v^2),                  
\label{2.19}
\ee
which is the metric of the Poincar\'e half-plane~\cite{GroSte87}.

\sect{The $h$-deformed Heisenberg algebra and \\
the Laplace-Beltrami operator }
                                                 
\subsect{ The $h$-deformed Heisenberg algebra}

In this Subsection, we shall construct a deformed Heisenberg algebra on the
extended $h$-deformed quantum plane. First, we shall
introduce $\pu, \pv $ such that 
\be 
\pu u = \pv v = 1.
\label{3.1}
\ee
We suppose the following ansatz
\be
\px = (\px u)\pu + (\px v)\pv, \hspace{1cm} 
 \py = (\py u)\pu + (\py v)\pv.
\label{3.2}
\ee
The coefficients can be calculated using Equation (\ref{2.5}), 
\be
\begin{array}{ll}
\px u = y^{-1}, & \px v = 0, \\
\py u = -xy^{-2}-hy^{-1}, & \py v = -2y^{-3}, 
\end{array}
\label{3.3}
\ee
and thus we have
\be
\pu = y\px, \hspace{1cm} 
\pv = -\frac{1}{2}(xy^2 - 2hy^3 )\px - \frac{1}{2}y^3 \py.
\label{3.4}
\ee
Then $\pu, \pv $ satisfy not only  Equation (\ref{3.1}) but also  
\be
 \pu v = \pv u = 0. 
\label{3.5}
\ee
Now that 
\be
\begin{array}{ll}
\pu x  = y, & \pu y = 0, \\
\pv x = -\frac{1}{2}(xy^2 - 2hy^3 ), & \pv y = -\frac{1}{2}y^3,
\end{array}
\label{3.6}
\ee
we can verify the relations
\be
\pu = (\pu x)\px + (\pu y)\py, \hspace{1cm}
\pv = (\pv x)\px + (\pv y)\py.
\label{3.7} 
\ee
Moreover, it is straightforward to see that
\be
d  = dx \partial_{x} + dy\partial_{y} = du \pu + dv \pv .
\label{3.8}
\ee

From Equations (\ref{2.5}) and (\ref{3.7}), it follows that
\be
\begin{array}{ll}
\, [u, \pu ] = -1 + 2h\pu, & [u, \pv ] = 2h \pv, \\
\, [v, \pu ] = 0, & [v, \pv ] = -1 \\
 \, [ \pu, \pv ]  = 0. &
\end{array}
\label{3.9}
\ee 
Now we introduce the  momentum operators  temporarily  by
\bea 
p_{u} = -i\hbar \pu, \hspace{1cm} p_{v} = -i\hbar \pv,
\label{3.10}
\eea
then we have the following commutation relations
\be
\begin{array}{ll}
 \,  [u, p_{u} ] = i \hbar + 2hp_{u}, & [u, p_{v} ] = 2hp_{v}, \\
 \,  [v, p_{u}] = 0,                 & [v, p_{v} ] = i\hbar, \\
 \,  [p_{u}, p_{v} ] = 0. &
\end{array}
\label{3.11}
\ee
These relations satisfy  
the Jacobi identities.
It is worth  noticing that the commutation relations are
 invariant under the transformation
\be
p_{u}' = p_u, \hspace{1cm} p_{v}' = p_v + \alpha v^{-1}
\label{3.12}
\ee
for any complex number $\alpha $.
Moreover, the Equation (\ref{3.11})  satisfies the hermiticity  if we
define the Hermitian adjoints of the momentum operators as follows:
\be
p_{u}^{\dagger} = p_u, \hspace{1cm} p_{v}^{\dagger} = p_v + \beta  v^{-1}
\label{3.13}
\ee
 for any complex number $\beta $. 

From now on, we choose   $\beta = 2i\hbar $  in Equation (\ref{3.13}). 
In fact, not only does this choice guarantee
the hermiticity of the Laplace-Beltrami operator
to be constructed in the next Subsection, but also
it is consistent with  that in
the commutative limit~\cite{GroSte87, Pod28}
once
one defines 
the innner product $< f, g >$ of two functions on  
the Poincar\'e half-plane  to be
$ < f, g> = \int  \bar{f} g d\mu $ 
with the measure $d\mu = \tilde{v}^{-2}d\tilde{u} d\tilde{v} $. 
It follows then that  
 $ p_{\tilde{v}}^{\dagger} = p_{\tilde{v}} + 2i\hbar \tilde{v}^{-1} $ 
and 
the Hermitian momentum operator is not $p_{\tilde{v}}$ but 
$P_{\tilde{v}} \equiv  p_{\tilde{v}} + i\hbar \tilde{v}^{-1}$.    
    Similarly, the  momentum operator is not $p_{v}$ but 
$P_{v} \equiv p_{v} + i \hbar v^{-1}$ in the quantum plane.
This   is not unusual  either in nonflat spaces~\cite{Kle95} or   
in quantum spaces~\cite{FicLorWes96}.
In fact, the second term of $P_{v}$ reflects the nonflatness
of the space.
 From Equation (\ref{3.12}), it follows that the Hermitian
 momentum operators $P_{u} \equiv p_{u} $
and $P_{v}$ satisfy the following $h$-deformed Heisenberg algebra  
on the $(u, v)$-quantum plane
which is of the same form
as in Equation (\ref{3.11}):
\be
\begin{array}{ll}
 \,  [u, P_{u} ] = i \hbar + 2hP_{u}, & [u, P_{v} ] = 2hP_{v}, \\
 \,  [v, P_{u}] = 0,                 & [v, P_{v} ] = i\hbar, \\
 \,  [P_{u}, P_{v} ] = 0. &
\end{array}
\label{3.16}
\ee

In this case, the Hermitian adjoints  
of $p_{x} $ and $p_{y} $ in Equation (\ref{2.50})
are given by
 \be
p_{x}^{\dagger} = p_x, \hspace{1cm} 
p_{y}^{\dagger} = p_y + 2hp_{x}. 
\label{3.14}
\ee           
and the operator such as $p_{y}$ is not Hermitian, 
in contrast to the assumption in \cite{Agh93}.                    
In fact, the Hermitian momentum operators are
   $P_x \equiv p_x $ and $P_y \equiv p_y + hp_x $.
Now we can write
 Equation (\ref{2.5}) as
 \be
\begin{array}{ll}
 \, [x, p_{x}] = i\hbar + hy p_{x}, 
     & [x, p_{y}] =  -hx p_{x} - h^2 y p_{x}
                  - h y p_{y}, \nonumber \\
 \,  [y, p_{x}] = 0,
      & [y, p_{y}] = i\hbar + hy p_{x}, \\
 \,  [p_{x}, p_{y}] =  h p^{2}_{x}.
 & \nonumber
\end{array}                                                  
 \label{3.114}
\ee
It is straightforward to see that the relations in the above Equation (\ref{3.114})
  satisfy
the hermiticity. Moreover, a lengthy calculation shows that they are
 covariant under the quantum group
$GL_{h}(2)$. 
Now  
 Equation (\ref{3.114}) yields immediately 
  the $h$-deformed Heisenberg algebra on the $(x, y)$-quantum plane
 \be
\begin{array}{ll}
 \, [x, P_{x}] = i\hbar + hy P_{x}, 
     & [x, P_{y}] =  i\hbar h -hx P_{x} + h^2 y P_{x}
                  - h y P_{y}, \nonumber \\
 \,  [y, P_{x}] = 0,
      & [y, P_{y}] = i\hbar + hy P_{x}, \\
 \,  [P_{x}, P_{y}] =  h P^{2}_{x}.
 & \nonumber
\end{array}                                                  
 \label{3.17}
\ee      
From the covariance of the Equation (\ref{3.114}) it follows easily that
this deformed Heisenberg algebra
 is also covariant under the quantum group
$GL_{h}(2)$. 
  The deformed Heisenberg algebra 
 is comparable with that in \cite{Agh93} which is
not covariant under the quantum group $GL_{h}(2)$.

\subsect{ The Laplace-Beltrami operator }

From now on,  we shall concern the $(u, v)$-plane.
From Equation (\ref{2.12}),  the differential algebra over ${\cal A}$
is the sum of ${\cal A}$ and the ${\cal A}$-modules 
$\Omega^1, \Omega^2 $ of 1-forms
and 2-forms: 
\be
 \Omega ({\cal A}) = {\cal A} \oplus \Omega^1 \oplus \Omega^2.
\label{4.1}
\ee
As in  the classical geometry,
 we can define the star operator $*$ on $\Omega ({\cal A}) $ 
with the metric given in Equation (\ref{metric}) by
\be
\begin{array}{ll} 
 \, *1 = \theta^1 \theta^2, &  *\theta^1 \theta^2  = 1,         \\
 \, *\theta^1 = \theta^2, &  *\theta^2 = -\theta^1.
\end{array}
\label{4.2}
\ee
Then for  $\theta \equiv -\lambda_{a}\theta^a $, we have 
$ * \theta = -\lambda_{1}\theta^2 + \lambda_{2}\theta^1 $  and as in classical geometry
the star operator $*$ satisfies
\be
** = (-1)^{p(2-p)},
\label{4.3}
\ee
where $p$ stands for the order of the form to be acted on.
Now we define $\delta : \Omega^p \rightarrow \Omega^{p-1} $
by
\be
\delta \varpi = (-1)^{2(p+1) + 1}*d*\varpi.
\label{4.4}
\ee
From Equation (\ref{2.13}), it  follows then that for $f \in {\cal A}$
\be
\begin{array}{ll}
\delta f = 0,   &   \delta (\theta^1 \theta^2 ) = 0, \\
 \delta \theta^1 = 0,  &
\delta \theta^2 = -1. 
\end{array}
\label{4.5}
\ee
Moreover, a straightforward calculation yields 
\be
\begin{array}{ll}
 \delta(f\theta^1\theta^2) = e_{2}f\theta^1
 - e_{1}f\theta^2,  &
 \delta \theta = \lambda_{2},  \\
 \delta(f\theta ) = (e_{1}f)\lambda_{1} 
 + (e_{2}f)\lambda_{2} + f\lambda_{2}, 
& \delta(\theta f ) = \lambda_{1} e_{1}f 
 + \lambda_{2}  e_{2}f + \lambda_{2} f. 
\end{array} 
\label{4.6}
\ee
If we define the Laplace-Beltrami operator $\triangle $
to be
\be
 - \triangle = \delta d + d\delta,
\label{4.7}
\ee
then  it is easy to see that for any  $f \in {\cal A}$ 
\be
\triangle f = e_{1}^2 f + e_{2}^{2} f + e_{2}f.
\label{4.8}
\ee                                
From Equation (\ref{2.18}), in the commutative limit,
it follows that 
$\triangle  $ goes over to 
\be
\tilde{\triangle} \equiv 
\tilde{v}^2(\partial_{\tilde{u}}^2 + \partial_{\tilde{v}}^2 ),
\label{4.9}
\ee
which is the Laplace-Beltrami operator on the Poincar\'{e} 
half-plane~\cite{GroSte87}.
Thus this result is consistent with that of Ref.~\cite{ChoMadPar98} that 
the extended $h$-deformed quantum plane is a noncommutative version
of the Poincar\'{e} half-plane, a surface of constant negative 
Gaussian curvature.

Now we claim that the operators $ e_{1} $ and 
$e_{2}$ are nothing but $v\pu $ and $-v\pv $ respectively.
In fact, the first claim 
 can be seen from mathematical induction together with
the following observations:
\be
\begin{array}{ll}
e_{1}u =  v\pu u, & e_{1}u^{-1} = v\pu u^{-1}, \\
e_{1}v =  v\pu v, & e_{1}v^{-1} =  v\pu v^{-1},
\end{array}
\label{4.10}
\ee 
and
\be
\begin{array}{ll}
e_{1}(uf) =  v\pu (uf), & e_{1}(u^{-1}f) = v\pu (u^{-1}f), \\
e_{1}(vf) =  v\pu (vf), & e_{1}(v^{-1}f) =   v\pu (v^{-1}f),
\end{array}
\label{4.11}
\ee 
where 
we have used the identity that
 $\pu f = u^{-1}f + u \pu (u^{-1}f) - 2h \pu ( u^{-1}f) $. 
Similarly, the second claim  can also be proved. 

Moreover, from the above claims it follows that
\be
\triangle = v^2 (\partial_{u}^2 + \partial_{v}^2 )
           = -\frac{1}{\hbar^2}v^2 (p_{u}^2 + p_{v}^2 ).
\label{4.12}
\ee
It is straightforward  to see that
$\triangle $ is Hermitian for the choice of $\beta = 2i\hbar $
in Equation (\ref{3.13}).
Moreover,
 the Laplace-Beltrami operator $\triangle $ can be written as
 \be
\triangle = \frac{1}{\sqrt{\mid g \mid } }
\partial_{\mu} \sqrt{\mid g \mid } g^{\mu \nu} \partial_{\nu },
\ee                                                                                
as expected, where
 $g^{\mu \nu } = g(du^{\mu } \otimes du^{\nu }) $ for $u^{\mu} = (u, v) $ 
and $\mid g \mid = \det (g_{\mu \nu})$.

\sect{ Quantum mechanics on the quantum plane }

\subsect{ Free particles }
                         
Now we shall discuss  quantum mechanics 
 on the $h$-deformed quantum plane.
We assume~\cite{Mad91, Pod28} that  the Schr\"{o}dinger equation of a particle with mass $m$
in a potential $V$ is given by 
\be
i\hbar \partial_{t} \Psi = \hat{H} \Psi ,
\ee                                         
where $\hat{H} = -\frac{\hbar^2}{2m}\triangle + V(u, v) $
and time $t$ is regarded as an extra commuting coordinate.
As in the commutative case, it is enough to find 
 the solutions of the time-independent Schr\"{o}dinger 
equation.

First let us solve the time-independent 
Schr\"odinger equation of a free particle of mass $m$ 
\be
-\frac{\hbar^2}{2m}\triangle \Psi (u, v) = E \Psi (u, v).
\label{5.1}
\ee               
Let us put
 $\Psi (u, v) = g(v)f(u) $ and $\lambda = - \frac{2mE}{\hbar^2} $. 
  Then we have 
\be
v^2 (\partial_{u}^2 + \partial_{v}^2 ) g(v)f(u) = \lambda g(v)f(u).
\label{5.2}
\ee
We can  decompose this equation into two equations without difficulty
using the relations in Equations (\ref{3.5}) and (\ref{3.9})
\be
  \partial_{u}^2 f(u) = -C^2f(u) 
\label{5.3-1}
\ee
and
\be
  v^2 \partial_{v}^2 g(v) = (C^2 v^2 + \lambda ) g(v),
\label{5.3-2}
\ee
where $C$ is a constant.
In the case when $ C = 0 $, it follows that  up to constant multiplication
$f = 1$  and for any number $\alpha \neq 0 $ 
\be
g(v) = v^{\alpha}
\ee
with $ E = -\frac{\hbar^2}{2m}\alpha (\alpha - 1) $. 
However, this solution is not normalizable in the commutative limit.
Thus we suppose that  $C \neq 0$.
From the commutation relations in Equation (\ref{3.9}),
it follows that for any constant $a$ 
\be
\partial_{u}e^{au} = \frac{1-e^{-2ah}}{2h}e^{au},
\label{5.4}
\ee                                                  
where $e^{au} = 1 + au + \frac{1}{2!}(au)^2 + \cdots $ is a formal
power series. 
Thus the solution of the  equation in (\ref{5.3-1}) is given by
\be
f(u) = e^{ iku}                                         
\label{5.5}
\ee                                                       
with
\be
C^2 = \left[\frac{1-e^{ -2i k h  }}{2\mid h \mid}\right]^2,
\label{5.6}
\ee                     
 which goes over to $k^2 $ in the commutative limit $h \rightarrow 0$.
On the other hand,
since the commutation relations between $v, \partial_{v}$ 
and $\tilde{v}, \partial_{\tilde{v}} $ are of  the same form,
we can regard  Equation (\ref{5.3-2}) as an ordinary differential equation
although the solutions of the quantum plane are  formal.
 Let  $\lambda = -\kappa^2 - \frac{1}{4} $. 
Now, if we put $ z = iCv $ 
and $g(v) = \sqrt{z}\phi (z) $,
then
the Equation (\ref{5.3-2})  becomes the Bessel differential equation  
\be
\phi''(z) + \frac{1}{z}\phi'(z) + (1 + \frac{\kappa^2}{z^2})\phi(z) = 0.
\label{5.8}
\ee                                                      
Thus the solution of the differential equation  
in (\ref{5.3-2})  is given by
\be
g(v) = \sqrt{v}K_{ i\kappa }(\mid C \mid v).
\ee        

              
The  energy eigenvalues
 $E = -\frac{\hbar^2}{2m} \lambda =  \frac{\hbar^2}{2m}(\kappa^2 + \frac{1}{4}) $
for $\kappa > 0 $ 
constitute  a continuous spectrum.
The largest lower bound state
with $\kappa = 0$ is not allowed since the normalized wave function
 vanishes identically in the commutative limit~\cite{GroSte87}. 
In the limit the Green function 
is known to have a cut on the positive real axis in the complex
energy plane with a branch point at $E = \frac{\hbar^2}{8m} $.
The Bessel functions in these solutions seem to be
 related to the `cylindrical' topology of the Poincar\'e
half-plane~\cite{Sti92}. 
  
For the classical mechanics, the equations of motion
with the Hamiltonian        
\be
H = \frac{\tilde{v}^2}{2m}(P_{\tilde{u}}^2 + P_{\tilde{v}}^2 ) 
\ee 
  yield the geodesic equation on the Poincar\'e half-plane:
\be
\ddot{\tilde{u}} = \frac{2\dot{\tilde{u}}\dot{\tilde{v}}}{\tilde{v}},
\hspace{1cm}                                      
\ddot{\tilde{v}} = 
\frac{1}{\tilde{v}}(\dot{\tilde{v}}^2 - \dot{\tilde{u}}^2 ).
\label{5.10}
\ee

We shall conclude this Subsection with the following observation.
As in the $q$-deformed case (see e.g.~\cite{SchWes92}), 
the first relation in Equation (\ref{3.9}) is the one whose Leibniz rule does
not involve the coordinate $v$.
Thus let us
assume a one-dimensional $h$-deformed Heisenberg relation in the $u$-direction
from Equation (\ref{3.16}) as follows:
\be
  [u, P_{u} ] = i\hbar + 2h P_{u}
\ee                              
and choose the Hamiltonian as
$ \hat{H} = \frac{1}{2m}P_{u}^{2} $.
Then 
the one-dimensional time-dependent Schr\"{o}dinger equation
has a solution of the form   
\be
\Psi (t, u) = e^{  i(ku - \omega t)}
\ee 
provided with
\be
\omega  = \frac{\hbar }{2m}\left[ \frac{1 -e^{ - 2ik  h }}{2\mid h \mid}\right]^{2}. 
\ee                                     
Thus the energy depends on the parameter $h$ explicitly, 
which is similar to the 1-dimensional
$q$-deformed case~\cite{CerHinMadWes98}.
 However, the $h$-dependence does not arise
in this way for a free particle on  the $h$-deformed quantum plane.
We investigate the $h$-dependence of the energy spectra 
for some bound states in the next Subsections.

\subsect{The motion under an oscillator-like potential }         

Let us consider a physical system with a Hamiltonian
\be
\hat{H} = -\frac{\hbar^2}{2m}v^{2}(\partial^2_{u} + \partial^2_{v}) 
  +v^2(A + \frac{1}{2}m\omega^2 v^2 )
\label{4-3.1}
\ee                                  
for some constants $A, \omega $.
The potential is sometimes called oscillator-like~\cite{Gro90}.
   As in the free particle case, we put 
\be
\hat{H}g(v)f(u) = Eg(v)f(u)
\label{4-3.2}
\ee                        
and decompose it into two differential equations 
\be
\partial^2_{u}f(u) = -C^2f(u)
\label{4-3.3}
\ee
and
\be
v^2\partial^2_{v}g(v) - (av^2 +  bv^4 + \lambda )g(v) = 0,
\label{4-3.4}
\ee
where 
\be
a = C^2 + \frac{2m}{\hbar^2}A, \qquad b = (\frac{m\omega}{\hbar})^2,
\qquad  \lambda = -\frac{2m}{\hbar^2}E.
\label{4-3.4-1}
\ee                 
The differential equation for $u$ in (\ref{4-3.3}) has a solution given by 
Equation (\ref{5.5}) with $C$ in  (\ref{5.6}).
If we put $ y = v^2 $, then from  Equation (\ref{4-3.4}) we obtain  
\be
4y^2\frac{d^2 g}{dy^2} + 2y\frac{dg}{dy} - (ay + by^2 + \lambda )g = 0.
\label{4-3.5}
\ee  
 Moreover, if we put $g = y^{-\frac{1}{4}}\phi (y) $, then
we have
\be
y^2 \frac{d^2 \phi (y)}{dy^2} 
- \frac{1}{4}( ay + by^2  + \lambda - \frac{3}{4})\phi (y) = 0.
\label{4-3.6}
\ee 
Now  let us write $\phi (y)$ as the following form 
\be
\phi (y) = y^{\beta}e^{-Ky}\xi (2Ky).
\label{4-3.7}
\ee                              
Then
Equation (\ref{4-3.6}) becomes
 a Laguerre  differential equation for $z = 2Ky $ 
\be
z\xi''(z) + (2\beta -z )\xi'(z) + (-\frac{a}{8K} - \beta )\xi(z) = 0
\label{4-3.8} 
\ee
 if $\beta $ and $K (K > 0)$ satisfy 
\be
\frac{1}{4}(\lambda - \frac{3}{4}) = \beta (\beta - 1), \qquad K^2 = \frac{b}{4}.
\label{4-3.9}
\ee
Thus bound state solutions are obtained 
by the associated Laguerre polynomials 
\be
\xi (z) = L_{n}^{(\nu )}(z)
\label{4-3.10} 
\ee
with                                          
\be
\nu + 1 = 2\beta > 0, \qquad n = -\frac{a}{8K} - \beta,
\label{4-3.11}
\ee
where $A$ should be negative enough such that $ a < 0$
 for the existence of bound states.
 If we put $V_{h} = \frac{\hbar^2}{2m}a $, that is 
\be
V_{h} =  -\frac{\hbar^2}{2m}
\left[\frac{1-e^{ -2i k h  }}{2h}\right]^2 + A,
\ee
then we have 
up to constant multiplication
\be
g(v) = v^{^{\textstyle \frac{\mid V_{h} \mid }{\hbar \omega } - 2n -\frac{1}{2}}  }
       e^{^{\textstyle -\frac{m\omega}{2\hbar} v^2 }     }
      L_{n}^{\textstyle 
    (\frac{\mid V_{h} \mid }{\hbar \omega } - 2n - 1)}(\frac{m \omega }{\hbar } v^2 )
\label{4-3.12}
\ee
with the energy eigenvalues
\be
E_{n} = \frac{\hbar^2}{8m} 
 - \frac{\hbar^2 }{2m}(\frac{\mid V_{h} \mid }{\hbar \omega } - 2n - 1)^2,
\label{4-3.13}
\ee
for $ n = 0, 1, \cdots , N_{M} < \frac{\mid V_{h} \mid }{2\hbar \omega } $.
The interpretation of this result in the commutative limit
is made in \cite{Gro90}.
On the $h$-deformed quantum plane,
not only the energy eigenvalues but also the number of bound states 
 depend explicitly on the deformation parameter $h$. 
In particular, we note that $a $ can not be less than $0$ when $k \rightarrow \infty $
in the commutative limit and thus  bound states can not exist.
However, for some $h$, $C^2 \rightarrow -\frac{1 }{4h^2} $ when  
 $k \rightarrow \infty $ and hence
bound states can still exist whenever such $h$ satisfies
\be
 A < \frac{\hbar^2 }{8m h^2 }.
\ee

\subsect{ The motion under a Coulomb-like potential }

Now we consider a particle
under a Coulom-like potential~\cite{Gro90} with a Hamiltonian
\be
\hat{H} = -\frac{\hbar^2}{2m}v^{2}(\partial^2_{u} + \partial^2_{v}) 
  +v^2(A + \frac{B}{2m}v^{-1} )
\label{4-4.1}
\ee                                  
for some constants $A $ and $ B $.
   As in the previous Subsection, we  decompose the Schr\"odinger
equation into two  
 differential equations 
\be
\partial^2_{u}f(u) = -C^2f(u)
\label{4-4.2}
\ee
and
\be
v^2\partial^2_{v}g(v) - (av + bv^2 + \lambda )g(v) = 0,
\label{4-4.3}
\ee
where 
\be
a = \frac{B}{\hbar^2}, \qquad b = C^2 + \frac{2m}{\hbar^2}A,
\qquad  \lambda = -\frac{2m}{\hbar^2}E.
\label{4-4.4}
\ee  
As before, we put  
\be
 g(v) = v^{\beta}e^{-Kv}\xi (2Kv).
\label{4-4.5}
\ee      
Then 
we have a Laguerre differential equation again for  $z = 2Kv $ 
\be
z\xi''(z) + (2\beta -z )\xi'(z) + (-\frac{a}{2K} - \beta )\xi(z) = 0
\label{4-4.6} 
\ee
if   $\beta $ and $K (K > 0)$ satisfy 
\be
\lambda  = \beta (\beta - 1), \qquad K^2 = b.
\label{4-4.7}
\ee
Thus  bound state solutions are also obtained  by        
the associated Laguerre polynomials
\be
\xi (z) = L_{n}^{(\nu )}(z)
\label{4-4.8} 
\ee
with                                          
\be
\nu + 1 = 2\beta > 0, \qquad n = -\frac{B}{2K\hbar^2} - \beta.
\label{4-4.9}
\ee
Here $ B $ should be less than $0 $ and $A$ should be such that $K > 0$ 
for the existence of the bound states.
 If we put $V_{h} = \frac{\hbar^2}{2m}C^2 + A $, then we have
up to constant multiplication
\be
g(v) = v^{^{\textstyle \frac{\mid B \mid }{2\hbar \sqrt{2mV_{h}} } - n}   }
       e^{^{\textstyle -\frac{\sqrt{2mV_{h}}}{\hbar} v }        }
      L_{n}^{\textstyle (\frac{\mid B \mid }{\hbar \sqrt{2mV_{h}}  } 
           - 2n - 1)}(2 \frac{\sqrt{2mV_{h}}}{\hbar} v )
\label{4-4.10}
\ee
with the energy eigenvalues
\be
E_{n} = \frac{\hbar^2}{8m} 
 - \frac{\hbar^2 }{2m}(\frac{\mid B \mid }{2\hbar \sqrt{2mV_{h}}  }
        - n - \frac{1}{2})^2,
\label{4-4.11}
\ee
for $ n = 0, 1, \cdots , N_{M} < \frac{\mid B \mid }{2\hbar \sqrt{2mV_{h}}  } $.
Not only the energy eigenvalues but also the number of bound states 
 depend on the deformation parameter $h$ as in the previous 
oscillator-like potential case. 
In particular,    no bound states can exist in the commutative limit
when $k \rightarrow \infty $
since $V_{h} \rightarrow \infty $.
However, 
bound states can still exist for some $h$ satisfying
\be
 A > \frac{\hbar^2 }{8m h^2 }.
\ee

\sect{ Conclusions }
 
  On the $h$-deformed quantum plane,
we have constructed the                           
$h$-deformed Heisenberg algebra   that
is covariant under the quantum group $GL_{h}(2)$.
It is worth  comparing it with 
the deformed Heisenberg algebra constructed by Aghamohammadi~\cite{Agh93}, 
 which is not covariant under the quantum group.
   We have also
constructed the Laplace-Beltrami operator on the quantum plane. This
operator has the Laplace-Betrami operator of the Poincar\'{e}
half-plane as its commutative limit. 

  We  have introduced
the Schr\"{o}dinger equations for a free particle and  particles under two specific
potentials  on
the quantum plane and find their solutions explicitly, taking values in 
the noncommutative algebra. 
In the commutative limit the behaviour of a quantum particle on
the quantum plane becomes that of the quantum particle on the
Poincar\'e half-plane, a surface of constant negative Gaussian
curvature.
 One usually expects that the energy spectra would depend on the parameter $h$
explicitly
at `large' momentum and this is the case for the two specific potentials in this work.
Moreover,   
it has been shown that bound states can survive on the quantum plane
 in a limiting case where 
bound states on the Poincar\'e half-plane disappear. 
 The bound state solutions for the two potentials are alike since the corresponding
Schr\"odinger equations can be reduced to the differential equation
for the confluent hypergeometric functions.

\section*{Acknowledgments}

This work was supported by Ministry of Education, Project No
BSRI-97-2414. 
The author would like to thank J Madore for a careful reading of this manuscript and 
 many fruitful suggestions.


\end{document}